\documentclass[12pt]{article}
\pdfoutput=1
\catcode`\@=11
\@addtoreset{equation}{section}

\global\arraycolsep=2pt 
\newcommand{\dd}{{\rm d}}

\newcommand{\ads}{\text{AdS} }

\newcommand{\ga}{\gamma}

\newcommand{\alg}[1]{\mathfrak{#1}}

\newcommand*{\AdS}[1]{\ensuremath{\text{AdS}_{#1}}}

\newcommand{\no}{\nonumber}
\newcommand{\mL}{\mathcal L}

\newcommand{\gP}{{\rm\bf P}}
\newcommand{\gJ}{{\rm\bf J}}
\newcommand{\gQ}{{\rm\bf Q}}

\usepackage{mathrsfs,amsbsy,amssymb,latexsym,amsfonts,amsmath,cite}
\usepackage{graphicx,color}
\usepackage{mathtools}

\RequirePackage[T1]{fontenc}
\RequirePackage[utf8]{inputenc}

\usepackage{physics}

\DeclareMathOperator{\STr}{STr}

\usepackage{enumitem}
\setitemize[1]{leftmargin=*}
\setenumerate[1]{leftmargin=*}

\allowdisplaybreaks[2]

\begin{document}

\begin{flushright}
\parbox{4cm}
{KUNS-2631
\\ \today
} 
\end{flushright}

\vspace*{1cm}

\begin{center}
{\Large \bf 
Generalized type IIB supergravity equations \\
and non-Abelian classical $r$-matrices
}
\vspace*{0.75cm} \\
{\large Domenico Orlando$^{\ast}$, 
Susanne Reffert$^{\ast}$, \\  
Jun-ichi Sakamoto$^{\dagger}$, 
and Kentaroh Yoshida$^{\dagger}$} 
\end{center}

\vspace*{0.25cm}

\begin{center}
$^{\ast}${\it Institute for Theoretical Physics, 
Albert Einstein Center for Fundamental Physics, 
University of Bern, Sidlerstrasse 5, CH-3012 Bern, 
Switzerland } 
\vspace*{0.25cm}\\
$^{\dagger}${\it Department of Physics, Kyoto University, \\ 
Kitashirakawa Oiwake-cho, Kyoto 606-8502, Japan} 
\end{center}

\vspace{1cm}

\begin{abstract}
We study Yang--Baxter deformations of the \(\AdS{5} \times S^5\) superstring 
with non-Abelian classical $r$-matrices which satisfy the homogeneous classical 
Yang--Baxter equation (CYBE). By performing a supercoset construction, 
we can get deformed \(\AdS{5} \times S^5\) backgrounds. 
While this is a new area of research, the current understanding 
is that Abelian classical $r$-matrices give rise to solutions of type IIB supergravity, 
while non-Abelian classical $r$-matrices lead to solutions of the generalized 
supergravity equations. We examine here some examples of non-Abelian classical 
$r$-matrices and derive the associated backgrounds explicitly. 
All of the resulting backgrounds satisfy the generalized equations. 
For some of them, we derive ``T-dualized'' backgrounds by adding a linear coordinate 
dependence to the dilaton and show that these satisfy the usual type IIB supergravity 
equations. Remarkably, some of the ``T-dualized'' backgrounds are locally identical to 
undeformed \(\AdS{5} \times S^5\) after an appropriate coordinate transformation, 
but this seems not to be generally the case. 
\end{abstract}

\setcounter{footnote}{0}
\setcounter{page}{0}
\thispagestyle{empty}

\newpage

\tableofcontents

\section{Introduction}

Yang--Baxter deformations provide a systematic way to study integrable deformations 
of non-linear sigma models in two dimensions. This method was invented by Klimcik 
for principal chiral models based on the modified classical Yang--Baxter equation 
(mCYBE)~\cite{Klimcik}. It was generalized to the symmetric coset case 
in~\cite{DMV} and the homogeneous CYBE case in~\cite{MY-YBE}.\footnote{
For the related affine algebras, see~\cite{KMY-alg,KOY,KY-Sch}}  

\medskip 

It is very interesting to consider some applications of Yang--Baxter deformations
in the context of the AdS/CFT correspondence~\cite{M}. 
The Green-Schwarz action of type IIB superstring on \(\AdS{5} \times S^5\) 
was constructed by Metsaev and Tseytlin in terms of the supercoset~\cite{MT}
\begin{equation}
 \frac{PSU(2,2|4)}{SO(1,4) \times SO(5)}\,. 
\end{equation}
This supercoset has a $\mathbb{Z}_4$-grading, 
generalizing the $\mathbb{Z}_2$-grading of the symmetric cosets. 
Hence the \(\AdS{5} \times S^5\) superstring exhibits classical integrability 
in the sense of kinematic integrability~\cite{BPR} 
(for nice reviews, see~\cite{AF-review,review,review1}).

\medskip 

The first application of Yang--Baxter deformation to the \(\AdS{5} \times S^5\) superstring 
was carried out by Delduc, Magro and Vicedo~\cite{DMV2}. The classical action was 
constructed with a classical $r$-matrix of Drinfeld-Jimbo type~\cite{DJ} and 
the associated symmetry algebra is a $q$-deformation of $\mathfrak{su}(2,2|4)$\,. 
This specific example is often called $\eta$-deformation. The metric and NS-NS two-form 
were subsequently computed by performing a coset construction for the bosonic element~\cite{ABF}. 
After that, some works were done towards constructing the full background 
by directly solving the equations of motion of type IIB supergravity~\cite{HRT,LRT}. 
This strategy was preferred simply because of the technical difficulty of the supercoset construction. 
In the end, Arutyunov, Borsato and Frolov accomplished doing the supercoset construction 
and the full background was determined~\cite{ABF2} (for a good review, see~\cite{Borsato}). 
A remarkable point is that the resulting background does not satisfy the equations of motion of 
type IIB supergravity. 

\medskip 

However, the situation is not hopeless. The ten-dimensional background is related to 
the full solution of the usual type IIB supergravity via T-dualities up to the linear dependence 
of the dilaton~\cite{HT-sol}. The world-sheet theory is not Weyl invariant 
but still scale invariant, \emph{i.e.}, the world-sheet beta function does not vanish 
but is given by a total derivative. From the viewpoint of the spacetime, 
this result indicates that the on-shell condition of type IIB supergravity is weakened to 
a generalized set of the equations. In fact, the generalized equations of type IIB supergravity 
were proposed in~\cite{scale} and the  background obtained in~\cite{ABF2} 
is a solution of the generalized equations. 
For a more general argument on the relation between the $\kappa$-symmetry 
of the $\eta$-deformed \(\AdS{5} \times S^5\) superstring and the generalized equations, 
see~\cite{WT}. 

\medskip 

Other examples of Yang--Baxter deformations of the \(\AdS{5} \times S^5\) superstring are 
based on the homogeneous CYBE~\cite{KMY-Jordanian-typeIIB}. In comparison to 
the mCYBE case, there are some advantages. One of them is that one can perform 
partial deformations of \(\AdS{5} \times S^5\)\,, i.e., only AdS$_5$ or only $S^5$ 
can be deformed. This class of Yang--Baxter deformations includes well-known examples 
such as gravity duals of non-commutative gauge theories~\cite{HI,MR}, 
the gamma-deformations of $S^5$~\cite{LM,Frolov}, and  Schr\"odinger 
spacetimes~\cite{MMT}\footnote{For the original works on Schr\"odinger geometries, see \cite{Orlando}}. 
In a series of works~\cite{LM-MY,MR-MY,Sch-MY,KMY-SUGRA, MY-duality,Stijn1,Stijn2}, 
the associated classical $r$-matrices were identified with these backgrounds\footnote{
For applications of Yang--Baxter deformation to other backgrounds such as AdS$_5\times T^{1,1}$\,, 
flat spacetime and pp-wave, see~\cite{CMY,YB-Min,YB-Min2,PvT,KY-NW}. 
For short summaries, see~\cite{MY-summary}. }. 
In a remarkable advance, the supercoset construction has been performed in~\cite{KY} 
and now the full background can be obtained for arbitrary classical $r$-matrices. 
For Abelian classical $r$-matrices, the R-R sector and dilaton have been confirmed 
for well-known backgrounds and it seems likely that Yang--Baxter deformations work 
well~\cite{KY}. However, for non-Abelian classical $r$-matrices, 
the associated backgrounds do not in general satisfy the equations of motion 
of type IIB supergravity~\cite{KY,HvT}. 

\medskip 

An interesting question here is whether deformed backgrounds 
for non-Abelian classical $r$-matrices satisfy the generalized equations of 
motion or not. {This question can be answered if the deformed string theory is supposed 
to be the canonical action of the Green-Schwarz string \cite{WT}. 
This assumption however has to be confirmed by a separate analysis and it is still an open problem. 
A confirmation was also given in~\cite{HvT} based on scaling limits of 
the $\eta$-deformed AdS$_5\times S^5$.} 
It may also be intriguing to study ``T-dualized'' backgrounds related to non-Abelian classical $r$-matrices 
and consider their  physical interpretation. We use quotation marks
for T-dualities involving solutions to the generalized SUGRA
equations, where the usual Buscher rules have to be supplemented with
a prescription for the behavior of the dilaton \cite{scale}.

\medskip 

In this paper, we consider Yang--Baxter deformations of the \(\AdS{5} \times S^5\) superstring 
with the homogeneous CYBE and focus on some examples of non-Abelian classical 
$r$-matrix. We derive the associated backgrounds explicitly and show that 
all of the resulting backgrounds satisfy the generalized equations. 
For some of them, ``T-dualized'' backgrounds are derived by adding a linear coordinate 
dependence to the dilaton and these satisfy the usual type IIB supergravity equations. 
Remarkably, some of the ``T-dualized'' backgrounds are locally T-dual to 
the undeformed \(\AdS{5} \times S^5\) via appropriate coordinate transformations,  
though it does not seem that this is always the case. 
At least in cases where the undeformed \(\AdS{5} \times S^5\) 
background can be reproduced, the classical integrability of the ``T-dualized'' background is manifest. 

\medskip 

This paper is organized as follows. Section~\ref{sec:yang-baxter-deformed} introduces 
Yang--Baxter deformations of 
the \(\AdS{5} \times S^5\) superstring based on the homogeneous CYBE 
and gives an outline of the supercoset construction. 
In Section~\ref{sec:generalized-type-iib}, we briefly introduce the generalized 
type IIB supergravity equations of motion.  
In Section~\ref{sec:examples}, we study six examples of non-Abelian classical $r$-matrices and 
the associated full backgrounds. For four of them, ``T-dualized'' backgrounds, which are solutions 
of the usual type IIB supergravity, are found. Then, three of the ``T-dualized'' backgrounds 
are shown to be locally equivalent to the undeformed \(\AdS{5} \times S^5\)\,. 
Section~\ref{sec:conclusions} is devoted to conclusion and discussion.

\section{Yang--Baxter deformed $\text{AdS}_5 \times S^5 $ backgrounds}
\label{sec:yang-baxter-deformed}

In this section, we shall give a short summary of Yang--Baxter deformations 
of the \(\AdS{5} \times S^5\) superstring based on the homogeneous CYBE 
and an outline of the supercoset construction. 

\subsection{The deformed string action}

The action of the deformed system is given by
\begin{eqnarray}
S=-\frac{\sqrt{\lambda_{\rm c}}}{4}\int_{-\infty}^\infty \dd{\tau}\int_0^{2\pi}\dd{\sigma}
\qty( \ga^{ab} - \epsilon^{ab})\,
\STr \qty[A_a\, d\circ\frac{1}{1-\eta R_g\circ d}(A_b)]\,, 
\label{YBsM}
\end{eqnarray}
where the left-invariant one-form $A_a$ is defined as
\begin{eqnarray}
A_a \equiv -g^{-1}\partial_a g\,,\quad\quad g\in SU(2,2|4) 
\end{eqnarray}
with the world-sheet index $a=(\tau,\sigma)$\,. 
We are in conformal gauge and the world-sheet 
metric takes the diagonal form $\ga^{ab}={\rm diag}(-1,+1)$\,. 
Hence there is no coupling of the dilaton to the world-sheet scalar curvature. 
The anti-symmetric tensor $\epsilon^{ab}$ is normalized as $\epsilon^{\tau\sigma}=+1$\,. 
The constant $\lambda_{\rm c}$ is the usual 't Hooft coupling.  
The deformation is measured by a constant parameter $\eta$\,. 
When $\eta=0$\,, the undeformed \(\AdS{5} \times S^5\) action~\cite{MT} is reproduced. 

\medskip

An important quantity here is the chain of operations $R_g$ defined as
\begin{eqnarray}
R_g(X)\equiv g^{-1}R(gXg^{-1})g\,, \quad\quad \forall X\in \alg{su}(2,2|4)\,,
\label{R}
\end{eqnarray}
where the linear operator $R:\alg{su}(2,2|4)\to \alg{su}(2,2|4)$ 
is a solution of the homogeneous CYBE: 
\begin{eqnarray}
  \comm{R(X)}{R(Y)}-R(\comm{R(X)}{Y} + \comm{X}{R(Y)})=0\,. \label{CYBE}
\end{eqnarray}
This $R$-operator is related to the {\it skew-symmetric} classical $r$-matrix 
in the tensorial notation through the following formula: 
\begin{eqnarray}
  R(X) = \STr_2[r(1\otimes X)] = \sum_i\qty( a_i \STr[b_iX] 
- b_i \STr[a_iX] )\,.
\label{R-r}
\end{eqnarray}
Here $r$ is written as 
\begin{eqnarray}
  r = \sum_ia_i \wedge  b_i\equiv\sum_i \qty(a_i \otimes b_i - b_i \otimes a_i )\qquad 
\mbox{with}\qquad a_i,~b_i\in\mathfrak{su}(2,2|4)\,.
\end{eqnarray}

\medskip

The projection operator $d$ is a linear combination of projectors defined as 
\begin{eqnarray}
d &\equiv& P_1+2P_2-P_3\,,  
\label{op-d}
\end{eqnarray}
where $P_{\ell}~(\ell=0,1,2,3)$ are projectors to the $\mathbb{Z}_4$-graded components 
of $\mathfrak{su}(2,2|4)$\,. In particular, $P_0(\mathfrak{su}(2,2|4))$ is 
a local symmetry of the classical action, $\mathfrak{so}(1,4)\oplus\mathfrak{so}(5)$\,.
The numerical coefficients in (\ref{op-d}) are fixed 
by requiring kappa-symmetry~\cite{MT,KMY-Jordanian-typeIIB}.

\subsection{A parametrization of the group element}

The classical action (\ref{YBsM}) is written in terms of a group
element $g \in SU(2,2|4)$ and a coordinate system can be introduced 
by fixing an explicit parametrization. 

\medskip 

First of all, the group element $g$ is represented by the product of 
the bosonic and fermionic elements $g_{\rm b}$ and $g_{\rm f}$ as follows:  
\begin{eqnarray}
g=g_{\rm b}\,g_{\rm f}\,. \label{para1}
\end{eqnarray}
For our later convenience, let us parametrize 
the bosonic element $g_{\rm b}$ as 
\begin{equation}
  \begin{aligned}
    g_{\rm b} & = g_{\rm b}{}^{\text{AdS}_5}\,g_{\rm b}{}^{\text{S}^5}\,,\\
    g_{\rm b}{}^{\text{AdS}_5} &=
    \exp[x^0\,P_0+x^1\,P_1+x^2\,P_2+x^3\,P_3 ]\,
    \exp[\qty(\log z) D] \,, \\
    g_{\rm b}{}^{\text{S}^5}&= \exp[\frac{i}{2}( \phi_1\,
    h_1+\phi_2\,h_2+\phi_3\,h_3)]\, \exp[\,\xi\,\gJ_{68}]
    \exp[-i\,r\,\gP_6]\,.
  \end{aligned}
\end{equation}
Here, $P_\mu~(\mu=0,\ldots,3)$ and $D$ are translations and dilatation $D$ 
in the four-dimensional conformal algebra $\mathfrak{su}(2,2)$, and $h_i~(i=1,2,3)$ are 
the Cartan generators of $\mathfrak{su}(4)$. For the other generators 
and the details of the notation, see~\cite{KY}. 
The coordinates $x^\mu$ and $z$ describe the Poincar\'e AdS$_5$\,,  
and $r\,,~\xi\,,\phi_i$ parametrize the round five-sphere.
The resulting metric is given by \(\dd{s^2} = \dd{s^2}_{\rm AdS_5} + \dd{s^2}_{\rm
  S^5}\), where
\begin{align}
   \dd{s^2}_{\rm AS_5} &= \frac{-(\dd{x^0})^2 +(\dd{x^1})^2 +(\dd{x^2})^2 + (\dd{x^3})^2}{z^2} 
 +\frac{\dd{z^2}}{z^2}\,,  \label{AdS5} \\ 
  \dd{s^2}_{\rm S^5} &= \dd{r^2} + \sin^2 r \dd{\xi^2} + \cos^2\xi \sin^2 r \dd{\phi_1^2} 
+ \sin^2r\sin^2\xi \dd{\phi_2^2} + \cos^2 r \dd{\phi_3^2}\,.
\label{S5}
\end{align}

\medskip 

Then the fermionic group element $g_{\rm f}$ is generated by the supercharges $\gQ^I$ 
as follows:  
\begin{eqnarray}
g_{\rm f} &=& \exp (\gQ^I {\theta_I})\,,  
\qquad \gQ^I \theta_I \equiv (\gQ^{\check{\alpha}\hat{\alpha}})^I\,
(\theta_{\check{\alpha}\hat{\alpha}})_I \quad 
(I=1,2;~\check{\alpha},\hat{\alpha}=1,\ldots,4)\,. 
\end{eqnarray}
Here, $\theta_I=(\theta_{\check{\alpha}\hat{\alpha}})_I$ are Grassmann-odd coordinates 
and correspond to a couple of 16-component Majorana-Weyl spinors satisfying 
the Majorana condition (see~\cite{KY} for details). 

\medskip 

The coset construction for the bosonic element $g_{\rm b}$ is relatively straightforward 
and the metric and NS-NS two-form of the corresponding deformed geometries  
are well-studied. However, the supercoset construction including 
the fermionic element $g_{\rm f}$ is quite complicated and until very recently,
the R-R sector and dilaton have not been studied.

\medskip 

Eventually, the supercoset construction has been carried out concretely 
for the $\eta$-deformed \(\AdS{5} \times S^5\) superstring~\cite{ABF2}. 
Following this pioneering work, it has been generalized to the homogeneous CYBE case~\cite{KY}. 
In the following, we will give an outline of the supercoset construction.  

\subsection{An outline of the supercoset construction}

In order to understand the geometries corresponding to the integrable
deformations, we need to express the deformed action in
Eq.(\ref{YBsM}) in terms of the component fields of type IIB
supergravity. In other words, we need to put it in the canonical form\footnote{Note here that 
it is assumed that the deformed system can be regarded as the canonical Green-Schwarz string. 
Rigorously speaking, this identification has to be confirmed by other arguments.}
\begin{multline}
S= -\frac{\sqrt{\lambda_{\rm c}}}{4}\int_{-\infty}^\infty\! \dd{\tau} 
\int_0^{2\pi}\! \dd{\sigma}\,\qty[
\gamma^{ab}\widetilde{G}_{MN}\partial_a X^M \partial_b X^N 
- \epsilon^{ab} B_{MN} \partial_a X^M \partial_b X^N
]  \\ 
 -\frac{\sqrt{\lambda_c}}{2}\, 
i\bar{\Theta}_I \qty(\gamma^{ab}\delta^{IJ}-\epsilon^{ab}\sigma_3^{IJ})\, 
\tilde{e}_a^m \Gamma_m \widetilde{D}^{JK}_{b}\Theta_K + \order{\theta^4}\,,  
\end{multline}
where the spinorial covariant derivative $\widetilde{D}$ is defined as ~\cite{CLPS}
\begin{multline}
\widetilde{D}^{IJ}_{a} \equiv \delta^{IJ}\qty(\partial_a
-\frac{1}{4}\tilde{\omega}_a^{mn}\Gamma_{mn} ) 
+\frac{1}{8}\sigma_3^{IJ}\tilde{e}^m_a H_{mnp} \Gamma^{np} \\
 -\frac{1}{8}{\rm e}^{\Phi} \qty[\epsilon^{IJ} \Gamma^p F_p 
+ \frac{1}{3!}\sigma_1^{IJ}\Gamma^{pqr}F_{pqr} 
+ \frac{1}{2\cdot 5!}\epsilon^{IJ}\Gamma^{pqrst}F_{pqrst} ]
\tilde{e}^m_a \Gamma_m\,. \label{canonical}
\end{multline}
Here, $\tilde{e}_M^m$ is the vielbein for the deformed metric, 
and $\Theta$ is a 32-component spinor composed of $\theta_I$. 

\medskip 

Although it is quite difficult to rewrite the action at all orders in $\theta$, 
the second-order result is enough for our purposes. 
This expansion simplifies the analysis that remains nevertheless quite
involved. For the technical details, see~\cite{ABF2,KY}.

\section{The generalized type IIB supergravity equations}
\label{sec:generalized-type-iib}

One of the main problems in the study of integrable deformations is to
determine if the corresponding ten-dimensional field content can be
understood as a string theory background.

\medskip

In the case of the \(\eta\)-deformation, the ten-dimensional
background can be obtained via the supercoset construction outlined
above and one can see that the fields \emph{do not satisfy} the type
IIB equations of motion~\cite{ABF2}. In fact, as was shown
in~\cite{scale}, the worldsheet theory is \emph{not Weyl invariant},
but only scale invariant: the corresponding beta function does not
vanish but is a total derivative.
This fact weakens the usual on-shell construction of type IIB
supergravity and has lead to the definition of \emph{generalized} type IIB
equations~\cite{scale}\footnote{For an earlier argument on the generalized equations 
at the linear order level, see~\cite{Mikhailov}.}. 
Note that as of now, this is a purely on-shell
construction for which no ten-dimensional effective action has been found yet. 

\medskip 

Interestingly, in  recent work~\cite{WT}, 
it has been shown that the $\kappa$-symmetry of the Green-Schwarz superstring generally 
leads to the generalized equations of type IIB supergravity, while in the older literature~\cite{old},  
it was shown that the on-shell condition of the usual type IIB supergravity leads to the kappa-invariance 
of the Green-Schwarz string action. 

\medskip

The generalized type IIB supergravity equations~\cite{scale} are
\begin{align}
  &R_{MN}-\frac{1}{4}H_{MKL}H_N{}^{KL}-T_{MN}+D_MX_N+D_NX_M=0\,, \label{general1} \\
  &\frac{1}{2}D^K H_{KMN}+\frac{1}{2}F^KF_{KMN}+\frac{1}{12}F_{MNKLP}F^{KLP} = X^K H_{KMN}+D_M X_N-D_N X_M\,, \label{general2}\\
  &R-\frac{1}{12}H^2+4D_MX^M-4X_MX^M=0\,,  \label{general3}\\
  &D^M\mathcal{F}_M-Z^M \mathcal{F}_M-\frac{1}{6}H^{MNK}\mathcal{F}_{MNK}=0\,, 
    \qquad I^M\mathcal{F}_M=0\,, \label{general4} \\
  &D^{K}\mathcal{F}_{KMN}-Z^K \mathcal{F}_{KMN} 
    - \frac{1}{6}H^{KPQ}\mathcal{F}_{KPQMN}
    - \qty(I\wedge \mathcal{F}_1)_{MN}=0\,, \label{general5} \\
  &D^{K}\mathcal{F}_{KMNPQ}-Z^K \mathcal{F}_{KMNPQ} +\frac{1}{36}\epsilon_{MNPQRSTUVW}H^{RST}\mathcal{F}^{UVW}
    - \qty(I\wedge \mathcal{F}_3)_{MNPQ}=0\,. 
\label{general6}
\end{align}
Here $M,N= 0, 1, \dots, 9$. The meaning of these equations is summarized as follows:
\begin{itemize}
\item The first equation (\ref{general1}) is for the metric 
in the string frame $G_{MN}$. The matter contribution $T_{MN}$ is given by 
\begin{multline}
  T_{MN} \equiv\frac{1}{2}\mathcal{F}_M\mathcal{F}_N
  +\frac{1}{4}\mathcal{F}_{MKL}\mathcal{F}_N{}^{KL}
  +\frac{1}{4\times 4!}\mathcal{F}_{MPQRS}\mathcal{F}_N{}^{PQRS} \\
  -\frac{1}{4}G_{MN} \qty(\mathcal{F}_K\mathcal{F}^{K}
  +\frac{1}{6}\mathcal{F}_{PQR}\mathcal{F}^{PQR})\,.
\end{multline}
Here $\mathcal{F}_M\,,\mathcal{F}_{MNK}\,,\mathcal{F}_{MNKPQ}$ are 
the rescaled R-R field strengths 
\begin{equation}
  \mathcal{F}_{n_1n_2\ldots}={\rm e}^{\Phi}F_{n_1n_2\ldots}\,,
\end{equation}
where $\Phi$ is the dilaton whose motion is described by
(\ref{general3})\,.
\item The second equation (\ref{general2}) is for the field strength
  $H_{MNK}$ of the NS-NS two-form. 
\item The fourth, fifth and 
sixth equations (\ref{general4}), (\ref{general5}) and (\ref{general6}) are 
for the R-R one-form, three-form and five-form field strengths.  
\end{itemize}

The Bianchi identities for the R-R field strengths are also generalized as 
\begin{align}
  &\qty(\dd{\mathcal{F}_1} - Z\wedge \mathcal{F}_1)_{MN} - I^K\mathcal{F}_{MNK}=0\,, \\
  &\qty(\dd{\mathcal{F}_3} - Z\wedge \mathcal{F}_3 + H_3\wedge \mathcal{F}_1)_{MNPQ} 
- I^K\mathcal{F}_{MNPQK} = 0\,, \\
  &\qty(\dd{\mathcal{F}_5} - Z\wedge \mathcal{F}_5 + H_3\wedge \mathcal{F}_3)_{MNPQRS} 
+ \frac{1}{6}\epsilon_{MNPQRSTUVW}I^T\mathcal{F}^{UVW} = 0\,. 
\end{align} 

\medskip 

Together with the standard type IIB fields, the equations (\ref{general1})-(\ref{general6}) 
involve the three new vector fields $X$, $I$ and $Z$. Let us consider them in detail.
In fact, only two of them are independent as the vector $X$ is expressed as  
\begin{eqnarray}
X_M\equiv I_M+Z_M\,.
\end{eqnarray}
$I$ and $Z$ satisfy the following relations: 
\begin{eqnarray}
D_M I_N+D_N I_M=0\,,\qquad D_M I_N-D_N I_M+I^K H_{KMN}=0\,, \qquad I^M Z_M=0\,.\label{IZ}
\end{eqnarray}
The first equation of (\ref{IZ}) is the Killing vector equation.
Assuming that $I_M$ is chosen such that the Lie derivative vanishes,
\begin{equation}
  (\mL_I B)_{MN}=I^K\partial_K B_{MN}+B_{KN}\partial_{M}I^K-B_{KM}\partial_N I^K=0,
\end{equation}
 the second equation of (\ref{IZ}) can be solved by
\begin{eqnarray}
Z_M=\partial_M\Phi-B_{MN}I^N\,.
\end{eqnarray}
Thus $Z$ can be regarded as a generalization of the dilaton gradient $\partial_M\Phi$\,.
In particular, when $I$ vanishes, $Z_M$ becomes $\partial_M\Phi$ and 
the generalized equations (\ref{general1})-(\ref{general6}) are reduced  
to the usual type IIB supergravity equations. 

\medskip 

As mentioned at the beginning of this section, the generalized equations 
(\ref{general1})-(\ref{general6}) were found in the study of the $\eta$-deformed 
\(\AdS{5} \times S^5\) superstring. However, they also appear for
Yang--Baxter deformations with \emph{non-Abelian} classical $r$-matrices satisfying 
the homogeneous CYBE\footnote{The $\eta$-deformation is an example of Yang--Baxter deformations 
with a non-Abelian classical $r$-matrix (the one of Drinfeld--Jimbo type) satisfying the mCYBE.}, 
as we will show in the next section.

\section{Examples of non-Abelian classical $r$-matrices}
\label{sec:examples}

In this section, we will consider six examples of non-Abelian classical $r$-matrices 
satisfying the homogeneous CYBE. The supercoset construction is performed explicitly 
by following~\cite{KY} and it is shown that all of the resulting backgrounds satisfy 
the generalized type IIB supergravity equations proposed in~\cite{scale}. 
For four of the examples, we derive ``T-dualized'' backgrounds 
which are solutions of the usual type IIB supergravity. 
Then for three of them, we show that the backgrounds are T-dual to the undeformed \(\AdS{5} \times S^5\) 
after an appropriate coordinate transformation. 

\medskip 

We will concentrate on deformations of the AdS$_5$ part and 
basically follow the notation and conventions for the $\mathfrak{su}(2,2)$ generators 
in~\cite{KY} to represent classical $r$-matrices.

\subsection{$r=P_1\wedge D$}

As a first example, let us consider the following non-Abelian classical $r$-matrix: 
\begin{equation}
  \label{eq:space-r-matrix}
r=\frac{1}{2}P_1\wedge D\,.
\end{equation}
Here $P_\mu$, $\mu=0,1,2,3$ are the generators of translations in the four-dimensional Poincar\'e algebra. 
The generator $D$ represents the dilatation. 
This is a solution of the homogeneous CYBE which was already used to study 
a Yang--Baxter deformation of four-dimensional Minkowski spacetime~\cite{YB-Min}. 

\medskip 

By performing the supercoset construction~\cite{KY}, 
the associated background is found to be\footnote{The metric and NS-NS two-form 
were computed in~\cite{Stijn2}.} 
\begin{equation}
  \begin{aligned}
    \dd{s}^2 &= \frac{z^2[\dd{t}^2+(\dd{x^1})^2+\dd{z}^2]+\eta^2(\dd{t}-t
      z^{-1}\dd{z})^2}{z^4+\eta^2(z^2+t^2)}
    +\frac{t^2(-\dd{\phi}^2+\cosh^2\phi \dd{\theta}^2)}{z^2}
    +\dd{s_{{\rm S}^5}^2}\,, \\
    B_2 &=\eta\,\frac{t \dd{t} \wedge \dd{x^1} + z \dd{z}\wedge \dd{x^1}}{z^4+\eta^2(t^2+z^2)}\,, \\
    F_3 &= -\frac{4\eta\, t^2\cosh\phi}{z^4}\left[\dd{t} \wedge
      \dd{\theta} \wedge \dd{\phi}
      -\frac{t}{z}\dd{\theta}\wedge \dd{\phi} \wedge \dd{z}  \right]\,, \\
    F_5 &=
    4\left[\frac{z^4}{z^4+\eta^2(t^2+z^2)}\omega_{\AdS5}+\omega_{{\rm
          S}^5}\right]\,,\\
    \Phi &= \frac{1}{2}\log
    \left[\frac{z^4}{z^4+\eta^2(t^2+z^2)}\right]\,,
  \end{aligned}
\label{space}
\end{equation}
where we have rewritten the four-dimensional Cartesian coordinates as:
\begin{align}
  \label{eq:cartesian-coord-x0x2x3}
  x^0 =t\sinh\phi\,,  \qquad x^2 =t\cosh\phi\cos\theta\,,  \qquad x^3 =t\cosh\phi\sin\theta\,.
\end{align}
Note here that the $\phi$ direction has the time-like signature. 
These fields \emph{do not} satisfy the equations of motion of type IIB
supergravity, but solve the generalized equations of
Section~\ref{sec:generalized-type-iib} when supplemented with the
following vectors:
\begin{align}
  I =\frac{\eta\, z^2}{z^4+\eta^2(t^2+z^2)} \dd{x^1}\,, \qquad 
 Z =-\frac{2\eta^2 t}{z^4+\eta^2(t^2+z^2)}\left(\dd{t}-\frac{t}{z}\dd{z}\right) \,.
\end{align}

\medskip

Let us now perform T-dualities for the deformed background
(\ref{space}). Following~\cite{scale}, the extra fields are traded for
a linear term in the dual dilaton. 
T-dualising along the $x^1$ and $\phi_3$ directions, we find:
\begin{equation}
  \label{T-4.3}
  \begin{aligned}
   \dd{s}^2 ={}& z^2\,(\dd{x^1})^2+\frac{1}{z^2}\Bigl[ (\dd{t} - \eta t
    \dd{x^1})^2+(\dd{z}-\eta z \dd{x^1})^2
    - t^2\dd{\phi}^2+t^2\cosh^2\phi \dd{\theta}^2\Bigr] \\
    &+ \dd{r}^2 +\sin^2 r \dd{\xi}^2+\cos^2\xi \sin^2 r\,
    \dd{\phi_1}^2
    + \sin^2r\sin^2\xi \dd{\phi_2}^2 + \frac{ \dd{\phi_3}^2}{\cos^2r}\,,\\
    \mathcal{F}_5 ={}&
    \frac{4t^2\cosh\phi }{z^4\cos r} (\dd{t}-\eta t \dd{x^1})\wedge
    (\dd{z}-\eta z \dd{x^1})
    \wedge \dd{\theta}\wedge \dd{\phi}\wedge \dd{\phi_3}  \\
    & +
    2z\sin^3 r\sin2\xi \dd{x^1}\wedge \dd{r}\wedge \dd{\xi} \wedge
    \dd{\phi_1}\wedge \dd{\phi_2}
    \,, \\
    \Phi ={}& -\eta x^1+\log\left[\frac{z}{\cos
        r}\right]\,. 
  \end{aligned}
\end{equation}
Remarkably, this is a solution 
of the \emph{usual} type IIB supergravity equations rather than the generalized ones. 
Note, however, that the dilaton has acquired a linear dependence on \(x^1\). 
This means that \(\partial_1\) is not an isometry and, as it is, 
the background cannot be dualized in the directions \((x^1, \phi_3)\) 
to go back to the initial frame. This result is very similar to the fact that 
the Hoare--Tseytlin solution~\cite{HT-sol} is ``T-dual'' to the $\eta$-deformed background. 
Indeed, we have followed the same strategy as in~\cite{HT-sol}. 

\medskip

The ``T-dualized'' background in Eq.~\eqref{T-4.3} is a solution to the standard type
IIB equations and has a remarkable property: it is \emph{locally equivalent} to 
undeformed \(\AdS{5} \times S^5\). Let us first perform the following change of coordinates:
\begin{align}
  t = \eta \rho \tilde{x}^1\,, \qquad 
                                  z = \eta \tilde{z} \tilde{x}^1\,, \qquad 
                                       x^1 = \frac{1}{\eta}\log(\eta \tilde{x}^1)\,.
\end{align}
Note that the new coordinate system does {\it not} cover all
of spacetime: the new coordinate $\tilde{x}^1$ has to be restricted to be positive (negative) 
when $\eta>0$ ($\eta<0$)\,. The signature of $\eta$ is fixed when we have chosen the deformation. 
This change of coordinates achieves the following points:
\begin{itemize}
\item it diagonalizes the metric;
\item it absorbs the \(x^1\)-dependence of the dilaton into the
  \(\tilde z\) variable, such that \(\partial_1\) is now a symmetry of the
  full background.
\end{itemize}
Explicitly, we find
\begin{equation}
  \begin{aligned}
    \dd{s}^2 ={}&\tilde{z}^2\,(\dd{\tilde{x}^1})^2+\frac{1}{\tilde{z}^2}\Bigl[
    \dd{\rho}^2+\dd{\tilde{z}}^2
    -\rho^2\dd{\phi}^2+\rho^2\cosh^2\phi \dd{\theta}^2\Bigr] \\
    & +\dd{r}^2 +\sin^2 r \dd{\xi}^2+\cos^2\xi \sin^2 r\,
    \dd{\phi_1}^2
    + \sin^2r\sin^2\xi \dd{\phi_2}^2 + \frac{ \dd{\phi_3}^2}{\cos^2r}\,,\\
    \mathcal{F}_5 ={}&
    \frac{4\rho^2\cosh\phi }{\tilde{z}^4\cos r} \dd{\rho}\wedge
    \dd{\tilde{z}}
    \wedge \dd{\theta}\wedge \dd{\phi}\wedge \dd{\phi_3} \\
    & +
    2\tilde{z}\sin^3 r\sin2\xi \dd{\tilde{x}^1}\wedge \dd{r}\wedge
    \dd{\xi} \wedge \dd{\phi_1}\wedge \dd{\phi_2} \\
    \Phi ={}& \log\left[\frac{\tilde{z}}{\cos r}\right]\,.
  \end{aligned}
\end{equation}
Now we can perform again the two standard T-dualities along \(\tilde{x}^1\) and
\(\phi_3\) to find, as advertised above, \emph{undeformed} \(\AdS{5} \times S^5\).\footnote{The
  usual Poincaré coordinates are found using the same change of
  coordinates as in Eq.~\eqref{eq:cartesian-coord-x0x2x3}.}

\medskip 

Let us stop for a moment to summarize what we have done. We have
started with a Yang--Baxter deformation of \(\AdS{5}\) described by the
non-Abelian \(r\)-matrix (Eq.~\eqref{eq:space-r-matrix}). Using the
supercoset construction outlined in
Section~\ref{sec:yang-baxter-deformed}, we have found the corresponding
ten-dimensional background (Eq.~\eqref{space}) that \emph{does not
  satisfy} the usual type IIB equations but is a solution to the
generalized equations described in Section~\ref{sec:generalized-type-iib}. Then
we have ``T-dualized'' this background using the rules of~\cite{scale}
to find a new background (Eq.~\eqref{T-4.3}) which solves the \emph{standard}
supergravity equations, but whose dilaton depends linearly on one of
the T-dual variables. Finally,
we have observed that after a change of
variables, this last background is nothing else than the T-dual of the
standard undeformed \(\AdS{5} \times S^5\). 
This result points to the fact that the Yang--Baxter deformation with the
classical $r$-matrix in Eq.~\eqref{eq:space-r-matrix} 
can be interpreted as an integrable twist, just like in the case of Abelian classical $r$-matrices 
(see for example~\cite{Frolov,AAF,Benoit,KKSY}). It might also suggest a relation between a gravity dual 
of the omega-background \cite{omega} and a non-abelian classical $r$-matrix.

\subsection{$r=P_0\wedge D$}

Our next example is the classical $r$-matrix
\begin{equation}
r=\frac{1}{2}P_0\wedge D\,,
\end{equation}
where \(P_0\) is the generator of time translations. This is a solution of the homogeneous CYBE 
which was originally utilized to study 
a Yang--Baxter deformation of 4D Minkowski spacetime~\cite{YB-Min}. 

\medskip 

By performing the supercoset construction~\cite{KY}, the associated
background is found to be\footnote{This background was studied in~\cite{Stijn2}, 
but only the metric and NS-NS two-form were computed therein.}
\begin{equation}
  \label{time}
  \begin{aligned}
    \dd{s}^2 ={}&\frac{z^2[-(\dd{x^0})^2+\dd{z}^2+\dd{\rho}^2]-\eta^2(\dd{\rho}-\rho
      z^{-1}\dd{z})^2}{z^4-\eta^2(z^2+\rho^2)}
    +\frac{\rho^2(\dd{\theta}^2+\sin^2\theta \dd{\phi}^2)}{z^2}
    +\dd{s_{{\rm S}^5}^2}\,, \\
    B_2 ={}&\eta\,\frac{z \dd{x^0} \wedge \dd{z} + \rho \dd{x^0} \wedge \dd{\rho}}{z^4-\eta^2(z^2+\rho^2)}\,, \\
    F_3 ={}&\frac{4\eta}{z^4}\left[\rho^2\sin\theta \dd{\rho}\wedge d
      \theta\wedge \dd{\phi}
      -\frac{\rho^3\sin\theta}{z}\dd{\theta}\,\wedge \dd{\phi} \wedge \dd{z} \right]\,, \\
    F_5 ={}&4 \left[\frac{z^4}{z^4-\eta^2(z^2+\rho^2)}\,\omega_{\AdS5} +
      \omega_{{\rm S}^5}\right]\,,\\
    \Phi ={}& \frac{1}{2}\log
    \left[\frac{z^4}{z^4-\eta^2(z^2+\rho^2)}\right]\,,
  \end{aligned}
\end{equation}
where the Cartesian coordinates of the four-dimensional Minkowski spacetime are 
\begin{align}
  x^1 &=\rho\sin\theta \cos\phi\,,& x^2 &=\rho\sin\theta\sin\phi\,, & x^3 &=\rho\cos\theta\,.
\end{align}
In this case we see that with the following two vectors
\begin{align}
  I = -\frac{\eta z^2}{z^4-\eta^2(\rho^2+z^2)} \dd{x^0}\,, \qquad 
  Z= \frac{2\eta^2 \rho}{z^4-\eta^2(\rho^2+z^2)}\left(\dd{\rho}-\frac{\rho}{z}\dd{z}\right)\,,
\end{align}
the above background becomes a solution of the generalized supergravity equations.

\medskip

Our next task is to perform ``T-dualities'' for the background
(\ref{time}) in the $x^0$ and $\phi_3$ directions. 
We find a solution of the standard type IIB supergravity
equations\footnote{Having performed a time-like T-duality we
  necessarily find a purely imaginary five-form flux.} with a dilaton that depends linearly on \(x^0\): 
\begin{equation}
  \label{T-4.2}
  \begin{aligned}
    \dd{s}^2 ={}& -z^2\,(\dd{x^0})^2+\frac{1}{z^2}\Bigl[ (\dd{\rho}+\eta
    \rho \dd{x^0})^2+(\dd{z}+\eta z \dd{x^0})^2
    + \rho^2\dd{\theta}^2 + \rho^2\sin^2\theta \dd{\phi}^2\Bigr]  \\
    & + \dd{r}^2+\sin^2 r \dd{\xi}^2+\cos^2\xi \sin^2 r \dd{\phi_1}^2
    + \sin^2r\sin^2\xi \dd{\phi_2}^2+\frac{ \dd{\phi_3}^2}{\cos^2r}\,,\\
    \mathcal{F}_5 ={}& -\frac{4i\rho^2\sin\theta }{z^4\cos r}
    (\dd{\rho}+\eta\rho \dd{x^0})\wedge(\dd{z}+\eta z \dd{x^0})
    \wedge \dd{\theta}\wedge \dd{\phi}\wedge \dd{\phi_3} \\
    &+ 2i z\sin^3 r\sin2\xi\,
    \dd{x^0}\wedge \dd{r}\wedge \dd{\xi} \wedge \dd{\phi_1}\wedge \dd{\phi_2}\,, \\
    \Phi ={}& \eta\, x^0+\log\left[\frac{z}{\cos
        r}\right]\,. 
  \end{aligned}
\end{equation}

\medskip

Finally, let us show that the ``T-dualized'' background (\ref{T-4.2}) is again equivalent to 
the undeformed \(\AdS{5} \times S^5\). 
First of all, we perform the following coordinate transformations: 
\begin{align}
\rho = \eta \tilde{\rho} \tilde{x}^0\,, \qquad 
z = \eta \tilde{z} \tilde{x}^0\,, \qquad
x^0 = -\frac{1}{\eta}\log(\eta \tilde{x}^0)\,.
\end{align}
Note here that the new coordinate $\tilde{x}^0$ is restricted to be positive 
(negative) when $\eta > 0$ ($\eta <0$). Just like in the previous case,
the metric is diagonal and the dilaton does not depend anymore on
\(\tilde{x}^0\), such that \(\partial_0\) is an isometry:
\begin{equation}
  \begin{aligned}
    \dd{s}^2 ={}&
    -\tilde{z}^2\,(\dd{\tilde{x}^0})^2+\frac{1}{\tilde{z}^2}\Bigl[
    \dd{\tilde{\rho}}^2+\dd{\tilde{z}}^2
    + \tilde{\rho}^2\dd{\theta}^2 + \tilde{\rho}^2\sin^2\theta \dd{\phi}^2\Bigr] \\
    & + \dd{r}^2+\sin^2 r \dd{\xi}^2+\cos^2\xi \sin^2 r \dd{\phi_1}^2
    + \sin^2r\sin^2\xi \dd{\phi_2}^2+\frac{ \dd{\phi_3}^2}{\cos^2r}\,,\\
    \mathcal{F}_5 ={}& -\frac{4i\tilde{\rho}^2\sin\theta
    }{\tilde{z}^4\cos r}\, \dd{\tilde{\rho}}\wedge \dd{\tilde{z}}
    \wedge \dd{\theta}\wedge \dd{\phi}\wedge \dd{\phi_3} \\
    &+ 2i \tilde{z}\sin^3 r\sin2\xi\,
    \dd{\tilde{x}^0}\wedge \dd{r}\wedge \dd{\xi} \wedge \dd{\phi_1}\wedge \dd{\phi_2}\,, \\
    \Phi ={}& \log\left[\frac{\tilde{z}}{\cos r}\right]\,.
  \end{aligned}
\end{equation}
Again, by performing two T-dualities along the $\tilde{x}^0$ and $\phi_3$
directions, we go back to 
the undeformed \(\AdS{5} \times S^5\) background.

\subsection{$r=(P_0-P_3)\wedge (D-L_{03})$}

Let us consider now the classical \(r\)-matrix
\begin{equation}
  r = \frac{1}{2\sqrt{2}} \pqty{P_0-P_3}\wedge \qty(D-L_{03})\,,
  \label{r-4.1}
\end{equation}
where \(L_{03}\) is the generator of the Lorentz rotation in the plane
\((x^0, x^3)\).

\medskip 

Performing the supercoset construction~\cite{KY}, we obtain the corresponding
background:
\begin{equation}
  \label{KY-sol}
  \begin{aligned}
    \dd{s^2} &= \frac{-2\dd{x^+}\dd{x^-} + \dd{\rho^2} + \rho^2\dd{\theta^2} + \dd{z^2}}{z^2}
    -\eta^2\left[\frac{\rho^2}{z^6}+\frac{1}{z^4}\right](\dd{x^+})^2 + \dd{s_{S^5}^2}\,, \\
    B_2 &= \eta\qty[\frac{\rho \dd{x^+} \wedge \dd{\rho}}{z^4}+\frac{1}{z^3}\dd{x^+}\wedge \dd{z} ]\,, \\
    F_3 &= 4\eta \qty[\frac{\rho^2}{z^5} \dd{x^+} \wedge \dd{\theta} \wedge
      \dd{z}
      +\frac{\rho}{z^4} \dd{ x^+}\wedge \dd{\rho} \wedge \dd{\theta} ]\,, \\
    F_5 &= 4 \qty(\omega_{\AdS5}+\omega_{{\rm S}^5}) \,, \\
    \Phi &= \Phi_0 \text{ (constant),} 
  \end{aligned}
\end{equation}
where the Cartesian coordinates of the four-dimensional Minkowski
spacetime $x^{\mu}$ are
\begin{align}
  x^\pm =\frac{1}{\sqrt{2}} \qty(x^0\pm x^3)\,, \qquad 
 x^1 =\rho \cos\theta\,, \qquad x^2 &=\rho \sin\theta\,. 
\end{align}
This background is a solution of the generalized 
supergravity equations\footnote{
This result is also supported in~\cite{HvT} based on a scaling argument of the $\eta$-deformed 
AdS$_5\times S^5$. 
} when supplemented by the vectors $I$ and $Z$:
\begin{align}
I = I_M \dd{x}^M = -\frac{2\eta}{z^2} \dd{x^+}\,,  \qquad Z_M = 0 = B_{MN}I^N\,.
\end{align}

Let us perform four ``T-dualities'' along the $x^+\,,x^-\,,\phi_1$ and $\phi_2$ directions\footnote{
To perform the T-dualities in the two light-like directions one can equivalently pass to 
Cartesian coordinates \((x^0, x^3)\), T-dualize in these and finally introduce light-like combinations 
for the T-dual variables.}. 
The resulting background is given by 
\begin{equation}
  \label{T-4.1}
  \begin{aligned}
    \dd{s}^2 ={}& -2z^2\dd{x^+}\dd{x^-}
    +\frac{(\dd{\rho}-\eta \rho\dd{x^-})^2+\rho^2\dd{\theta}^2+(\dd{z}-\eta z\dd{x^-})^2}{z^2} \\
     &+ \dd{r}^2 + \sin^2 r\, \dd{\xi}^2+\frac{ \dd{\phi_1}^2}{\cos^2\xi
      \sin^2 r}
    +\frac{ \dd{\phi_2}^2}{\sin^2r\sin^2\xi}+\cos^2r \dd{\phi_3}^2\,,   \\
    \mathcal{F}_5 ={}& \frac{4 i \rho}{z^3\sin\xi \cos\xi\sin^2 r}
    (\dd{\rho}-\eta \rho\dd{x^-})\wedge \dd{\theta} \wedge (\dd{z}-\eta z\dd{x^-})
    \wedge \dd{\phi_1}\wedge \dd{\phi_2}  \\
    & + 4i z^2\sin r\cos r \dd{x^+}\wedge \dd{x^-}\wedge \dd{r} \wedge \dd{\xi}\wedge \dd{\phi_3}\,, \\
    \Phi ={}& -2\eta x^-+\log \qty[\frac{z^2}{\sin^2r
        \sin\xi\cos\xi}]\,,
  \end{aligned}
\end{equation}
where the other components are zero.

\medskip

The ``T-dualized'' background in Eq.~\eqref{T-4.1} is a solution to the standard type
IIB equations and is again  \emph{locally} equivalent to 
undeformed \(\AdS{5} \times S^5\). Let us first change the coordinates as follows:
\begin{align}
  x^- = \frac{1}{2 \eta} \log( \tilde{x}^-)\, , \qquad \rho &= \tilde \rho \sqrt{\tilde{x}^-}  \, , 
\qquad   z = \tilde z \sqrt{\tilde{x}^-} \, .
\end{align}
Explicitly, we find
\begin{equation}
  \label{eq:light-like-IIB-solution}
  \begin{aligned}
    \dd{s}^2 ={}& -2\tilde{z}^2\dd{x^+} \dd{\tilde{x}^-} +\frac{\dd{\tilde{\rho}}^2
      +\tilde{\rho}^2\dd{\theta}^2+\dd{\tilde{z}}^2}{\tilde{z}^2} \\
    &+ \dd{r}^2 + \sin^2 r \dd{\xi}^2+\frac{ \dd{\phi_1}^2}{\cos^2\xi \sin^2 r}
    +\frac{ \dd{\phi_2}^2}{\sin^2r\sin^2\xi}+\cos^2r \dd{\phi_3}^2\,,   \\
    \mathcal{F}_5 ={}& \frac{4 i \tilde{\rho}}{\tilde{z}^3\sin\xi
      \cos\xi\sin^2 r}\,
    \dd{\tilde{\rho}}\wedge \dd{\theta}\wedge \dd{\tilde{z}}\wedge \dd{\phi_1}\wedge \dd{\phi_2} \\
    &+ 4i \tilde{z}^2\sin r\cos r\,
    \dd{x^+}\wedge \dd{\tilde{x}^-}\wedge \dd{r} \wedge \dd{\xi}\wedge \dd{\phi_3}\,,  \\
    \Phi ={}& \log \qty[\frac{\tilde{z}^2}{\sin^2 r
        \sin\xi\cos\xi}]\,.
  \end{aligned}
\end{equation}
Now, rewriting the light-like coordinates in terms of the Cartesian
coordinates as
\begin{align}
  x^+ &\equiv \frac{1}{\sqrt{2}}(\tilde{x}^0 + \tilde{x}^3)\,, &
\tilde{x}^- &\equiv \frac{1}{\sqrt{2}}(\tilde{x}^0 - \tilde{x}^3) \,,
\end{align}
and performing four T-dualities along $\tilde{x}^0$, $\tilde{x}^3$, $\phi_1$ and $\phi_2$, 
we reproduce the \emph{undeformed} \(\AdS{5} \times S^5\) background.


\paragraph{Mixing of Abelian and non-Abelian classical $r$-matrices.}

This example admits a generalization, obtained by mixing Abelian and non-Abelian classical $r$-matrices:
\begin{equation}
  r = \frac{1}{2\sqrt{2}}(P_0-P_3)\wedge \left[a_1 (D-L_{03})-a_2 L_{12}\right]\,. 
\label{4.7-r}
\end{equation}
When $a_2=0 $, the classical $r$-matrix reduces to the one
described above; when $a_1=0$, the $r$-matrix becomes Abelian 
and the associated background is the Hubeny--Rangamani--Ross solution
of~\cite{HRR}, as shown in~\cite{KY}. 

\medskip 

In~\cite{KY} it was shown that with a supercoset construction, one
finds the following ten-dimensional background
\begin{equation}
  \begin{aligned}
    \dd{s}^2 &=
    \frac{-2\dd{x^+}\dd{x^-}+\dd{\rho}^2+\rho^2\dd{\theta}^2+\dd{z}^2}{z^2}
    -\eta^2\left[(a^2_1+a^2_2)\frac{\rho^2}{z^6}+\frac{a_1^2}{z^4}\right](\dd{x^+})^2+\dd{s}_{\rm S^5}^2\,, \\
    B_2 &= \frac{\eta}{z^4}\dd{x^+}\wedge \qty[ a_1 \qty( \rho \dd{\rho} + z \dd{z} )
      - a_2 \rho^2 \dd{\theta} ]\,, \\
    F_3 &= \frac{4\eta\rho}{z^5}\dd{x^+}\wedge \qty[ a_1 \qty( z \dd{\rho} - \rho \dd{z} ) \wedge \dd{\theta}
      + a_2 \dd{\rho}\wedge \dd{z} ]\,, \\
    F_5 &= 4 \qty(\omega_{\AdS5} + \omega_{{\rm S}^5})\,, \\
    \Phi &= \Phi_0 \text{ (constant).}
  \end{aligned}
\end{equation}
This background is still a solution of the generalized equations with the two vectors $I$ and $Z$ given by
\begin{align}
  I = -\frac{2\eta a_1}{z^2}\dd{x^+}\,, \qquad Z_M &= 0\,.
\end{align}
In the special case $a_1=0$, the above background reduces to 
a solution of standard type IIB supergravity. 

\medskip

Let us next take four ``T-dualities'' along the $x^+\,,x^-\,,\phi_1$ and $\phi_2$ directions.
Then we can obtain a solution of the usual type IIB supergravity as 
\begin{equation}
  \label{T-4.7}
  \begin{aligned}
    \dd{s}^2 ={}&-2z^2\dd{x^+}\dd{x^-} +\frac{(\dd{\rho}-\eta a_1
      \rho\dd{x^-})^2+\rho^2(\dd{\theta}+\eta a_2\dd{x^-})^2
      +(\dd{z}-\eta a_1 z\dd{x^-})^2}{z^2} \\
    & + \dd{r}^2 + \sin^2 r \dd{\xi}^2+\frac{
      \dd{\phi_1}^2}{\cos^2\xi \sin^2 r}
    +\frac{ \dd{\phi_2}^2}{\sin^2r\sin^2\xi}+\cos^2r \dd{\phi_3}^2\,,  \\
    \mathcal{F}_5 ={}& \frac{4 i \rho}{z^3\sin\xi \cos\xi\sin^2 r}
    (\dd{\rho}-\eta a_1 \rho\dd{x^-})\wedge (\dd{\theta}+\eta
    a_2\dd{x^-})
    \wedge (\dd{z}-\eta a_1 z\dd{x^-})\wedge \dd{\phi_1}\wedge \dd{\phi_2}  \\
    & + 4i z^2\sin r\cos r \dd{x^+}\wedge \dd{x^-}\wedge \dd{r} \wedge \dd{\xi}\wedge \dd{\phi_3}\,,  \\
    \Phi ={}& -2\eta a_1x^-+\log\left[\frac{z^2}{\sin^2r
        \sin\xi\cos\xi}\right]\,, 
  \end{aligned}
\end{equation}
where the other components are zero.
It is easy to see that this is just a twist of the previous solution
(in Eq.~\eqref{eq:light-like-IIB-solution}) and in fact there is a
change of variables 
\begin{align}
  \rho &= \tilde{\rho}\,{\rm e}^{\eta a_1\, x^-}\,, &
                                                      z &= \tilde{z}\,{\rm e}^{\eta a_1\, x^-}\,, &
\theta &= \tilde{\theta}-\eta\, a_2 x^-\,, &
x^- &= \frac{1}{2\eta a_1}\log(2\eta a_1\, \tilde{x}^-)\,,
\end{align}
that maps this background to the same local form:
\begin{equation}
  \begin{aligned}
    \dd{s}^2 ={}& -2\tilde{z}^2\dd{x^+}d\tilde{x}^-+\frac{\dd{\tilde{\rho}}^2
      +\tilde{\rho}^2\dd{\tilde{\theta}}^2+\dd{\tilde{z}}^2}{\tilde{z}^2} \\
    & + \dd{r}^2 + \sin^2 r \dd{\xi}^2+\frac{
      \dd{\phi_1}^2}{\cos^2\xi \sin^2 r}
    +\frac{ \dd{\phi_2}^2}{\sin^2r\sin^2\xi}+\cos^2r \dd{\phi_3}^2\,,   \\
    \mathcal{F}_5 ={}& \frac{4 i \tilde{\rho}}{\tilde{z}^3\sin\xi
      \cos\xi\sin^2 r}\,
    \dd{\tilde{\rho}}\wedge \dd{\tilde{\theta}}\wedge \dd{\tilde{z}}\wedge \dd{\phi_1}\wedge \dd{\phi_2}  \\
    & + 4i \tilde{z}^2\sin r\cos r\,
    \dd{x^+}\wedge \dd{\tilde{x}^-}\wedge \dd{r} \wedge \dd{\xi}\wedge \dd{\phi_3}\,, \\
    \Phi ={}&\log\left[\frac{\tilde{z}^2}{\sin^2r
        \sin\xi\cos\xi}\right]\,,
  \end{aligned}
\end{equation}
which is a T-dual of the undeformed $\AdS{5}\times {\rm S}^5$ background.

\subsection{$r=(P_0-P_3)\wedge D$}

Let us now consider the non-Abelian classical $r$-matrix given by 
\begin{equation}
  r=\frac{1}{2\sqrt{2}} \qty( P_0 - P_3 ) \wedge D\,,
\end{equation}
which is another solution of the homogeneous CYBE. 

\medskip 

Using the supercoset construction~\cite{KY}, 
the associated background is found to be\footnote{The metric and NS-NS two-form were 
computed in~\cite{Stijn2}.} 
\begin{equation}
  \label{4.3}
  \begin{aligned}
    \dd{s}^2 ={}&\frac{1}{z^4-\eta^2(x^+)^2}\biggl[z^2(-2\dd{x^+}\dd{x^-}+\dd{z}^2)
    +2\eta^2z^{-2}x^+\rho \dd{x^+}\dd{\rho}-\eta^2 z^{-2}\rho^2(\dd{x^+})^2 \\
    &-\eta^2(\dd{x^+}-x^+z^{-1}\dd{z})^2\biggr]+\frac{\dd{\rho}^2+\rho^2\dd{\theta}^2}{z^2}
    +\dd{s_{{\rm S}^5}^2}\,, \\
    B_2 ={}& \eta\,\frac{\dd{x^+} \wedge (z\dd{z}+\rho \dd{\rho}-x^+\dd{x^-})}{z^4-\eta^2(x^+)^2}\,, \\
    F_3 ={}& 4\eta\,\frac{\rho}{z^4}\left[\frac{\rho}{z}\dd{x^+}\wedge \dd{\theta}\wedge \dd{z}
      + \dd{x^+}\wedge \dd{\rho}\wedge \dd{\theta}
      -\frac{ x^+}{z}\dd{\rho}\wedge \dd{\theta} \wedge \dd{z}\right]\,, \\
    F_5 ={}& 4\left[\frac{z^4}{z^4-\eta^2(x^+)^2}\,\omega_{\AdS5} + \omega_{{\rm S}^5}\right]\,, \\
    \Phi ={}& \frac{1}{2}\log \left[\frac{z^4}{z^4-\eta^2(x^+)^2}\right]\,. 
  \end{aligned}
\end{equation}
Here the following new coordinates have been introduced:
\begin{eqnarray}
x^0=\frac{x^++x^-}{\sqrt{2}}\,,\qquad
x^3=\frac{x^+-x^-}{\sqrt{2}}\,,\qquad
x^1=\rho\cos\theta\,,\qquad
x^2=\rho\sin\theta
\,.\label{lccoord}
\end{eqnarray}
This background satisfies the generalized SUGRA equations with the two vectors $I$ and $Z$ given by
\begin{align}
  I &=-\frac{\eta\, z^2}{z^4-\eta^2 (x^+)^2}\dd{x^+}\,, & Z & =\frac{2\eta^2x^+}{z^4-\eta^2 (x^+)^2}\dd{x^+}-\frac{2\eta^2(x^+)^2}{z(z^4-\eta^2 (x^+)^2)}\dd{z} \,.
\end{align}

As of now, we have not found an appropriate T-dual frame
in which this background is a solution to the standard type IIB
equations with a linear dilaton. It would be interesting to
understand if this means that we need a different realization
of a scale (non-Weyl) invariant worldsheet theory, different from the
one discussed in~\cite{scale}.

\subsection{The light-like $\kappa$-Poincar\'e $r$-matrix}

Let us next consider the following non-Abelian classical $r$-matrix: 
\begin{equation}
  r = \frac{1}{2\sqrt{2}}\Bigl[(L_{01}-L_{31})\wedge 
  P_1+(L_{02}-L_{32})\wedge P_2+L_{03}\wedge(-P_0+P_3)\Bigr]\,.
\end{equation}
This is a solution of the homogeneous CYBE which was employed 
to study a light-like deformation of the four-dimensional Poincar\'e algebra~\cite{LC-kappa}
and was also utilized to study a Yang--Baxter deformation of
four-dimensional Minkowski spacetime~\cite{YB-Min2}.

\medskip 

By performing the supercoset construction~\cite{KY}, the associated background is determined 
to be\footnote{The metric and NS-NS two form were computed in~\cite{Stijn2}. } 
\begin{equation}
  \begin{aligned}
    \dd{s}^2 ={}&\frac{1}{z^4-\eta^2(x^+)^2}\biggl[z^2(-2\dd{x^+}\dd{x^-}+\dd{z}^2)
    +2\eta^2z^{-2}x^+\rho \dd{x^+}\dd{\rho}-\eta^2 z^{-2}\rho^2(\dd{x^+})^2 \\
    &-\eta^2(x^+)^2z^{-2}\dd{z}^2\biggr]+\frac{\dd{\rho}^2+\rho^2\dd{\theta}^2}{z^2}
    +\dd{s_{{\rm S}^5}^2}\,, \\
    B_2 ={}& \eta\,\frac{\dd{x^+} \wedge (\rho \dd{\rho}-x^+\dd{x^-})}{z^4-\eta^2(x^+)^2}\,, \\
    F_3 ={}& 4\eta\,\frac{\rho}{z^5}\left(\rho \dd{x^+}\wedge
      \dd{\theta}\wedge \dd{z}
      - x^+\dd{\rho}\wedge \dd{\theta} \wedge \dd{z}\right)\,, \\
    F_5 ={}&4\left[\frac{z^4}{z^4-\eta^2(x^+)^2}\omega_{\AdS5}+\omega_{{\rm
          S}^5}\right]\,,\\
    \Phi ={}& \frac{1}{2}\log \left[\frac{z^4}{z^4-\eta^2(x^+)^2}\right]\,.
  \end{aligned}
\end{equation}
Here the coordinate system (\ref{lccoord}) has been used. This background does not 
satisfy the equations of the usual type IIB supergravity, but the generalized equations 
with the two vectors $I$ and $Z$ given by
\begin{align}
  I &=\frac{3\eta\, z^2}{z^4-\eta^2 (x^+)^2}\dd{x^+}\,, &
                                                          Z &=-\frac{2\eta^2x^+}{z^4-\eta^2 (x^+)^2}\dd{x^+}-\frac{2\eta^2(x^+)^2}{z(z^4-\eta^2 (x^+)^2)}\dd{z} \,.
\end{align}
This example is quite close to the one discussed in the previous
subsection. Also in this case, a T-duality to a solution of the standard type IIB equations is as of now missing.

\subsection{A scaling limit of the Drinfeld--Jimbo $r$-matrix 
\label{scaling:sec}}

Our last example is the  classical $r$-matrix
\begin{equation}
  r = -D\wedge P_0-L_{0\mu}\wedge P^{\mu}-L_{12}\wedge P_2-L_{13}\wedge P_3\,. \label{HvT-r}
\end{equation}
It was originally studied in~\cite{HvT} in relation to a scaling limit of the classical $r$-matrix 
of Drinfeld-Jimbo type. 

\medskip 

By performing the supercoset construction\footnote{
In this case, a non-trivial axion is turned on and hence the supercoset construction 
procedure in~\cite{KY} has to be extended by letting $U$ include 
$\lambda^{mn}\lambda^{pq}\Gamma_{mnpq}$\,. For details, see~\cite{future}. }\,, 
the full background is determined to be\footnote{The metric and NS-NS two-form 
were computed in~\cite{HvT} without the total derivative term in $B_2$.}
\begin{equation}
  \label{HvT}
  \begin{aligned}
    \dd{s}^2 &= \frac{-(\dd{x^0})^2+\dd{z}^2}{z^2-4\eta^2}
    +\frac{z^2\left[(\dd{x^1})^2+\dd{\rho}^2\right]}{z^4+4\eta^2\rho^2}
    +\frac{\rho^2\dd{\theta}^2}{z^2}+ \dd{s}^2_{{\rm S}^5}\,,  \\
    B_2 &= \frac{2\eta }{z(z^2-4\eta^2)} \dd{x^0}\wedge \dd{z}
    +\frac{2\eta\, \rho}{z^4+4\eta^2 \rho^2} \dd{x^1}\wedge \dd{\rho}\,, \\
    F_{1} &= \frac{16\eta^2\rho^2}{z^4}\dd{\theta}\,, \\
    F_{3} &= \frac{8\eta \rho^2}{z^3(z^2-4\eta^2)} \dd{x^0}\wedge
    \dd{\theta} \wedge \dd{z}
    +\frac{8\eta \rho}{z^4+4\eta^2\rho^2} \dd{x^1}\wedge \dd{\rho} \wedge \dd{\theta}\,,  \\
    F_{5} &=
    4\left[\frac{z^6}{(z^2-4\eta^2)(z^4+4\eta^2\rho^2)}\,\omega_{\AdS5}
      +\omega_{{\rm S}^5}\right] \,, \\
    \Phi &=
    \frac{1}{2}\log\left[\frac{z^6}{(z^2-4\eta^2)(z^4+4\eta^2\rho^2)}\right]\,.
  \end{aligned}
\end{equation}
Again, this background does not satisfy the usual type IIB supergravity equations, 
but becomes a solution of the generalized equations by taking $I$ and $Z$ as 
\begin{align}
I &= -\frac{8\eta \dd{x^0}}{z^2-4\eta^2}+\frac{4z^2\eta \dd{x^1}}{z^4+4\eta^2\rho^2}\,, &
Z &= \left[\frac{2(z^2+2\eta^2)}{z(z^2-4\eta^2)}-\frac{2z^3}{z^4+4\eta^2\rho^2}\right]\dd{z}
+\frac{4\eta^2\rho \dd{\rho}}{z^4+4\eta^2\rho^2}\,. \label{I-6}
\end{align}

\medskip

Performing ``T-dualities'' for all of the $U(1)$ directions, 
we can obtain a solution of the standard type IIB supergravity:  
\begin{equation}
  \begin{aligned}
    \dd{s}^2 ={}& -z^2\,(\dd{x^0})^2+z^2\,(\dd{x^1})^2
    +\frac{(\dd{\rho}+2\eta\rho \dd{x^1})^2+(\dd{z}+2\eta z
      \dd{x^0})^2}{z^2}
    +\frac{z^2\dd{\theta}^2}{\rho^2} \\
    & + \dd{r}^2+\sin^2 r \dd{\xi}^2+ \frac{\dd{\phi_1}^2}{ \sin^2
      r\cos^2\xi} + \frac{\dd{\phi_2}^2}{\sin^2r\sin^2\xi}
    +\frac{ \dd{\phi_3}^2}{\cos^2r}\,,\\
    \mathcal{F}_5 ={}&
    \frac{-4i}{z^2\sin^2r\cos r\sin\xi\cos\xi}\, \left(
      \dd{\rho}+2\eta\, \rho \dd{x^1}\right) \wedge
    \left(\dd{z}+2\eta\, z \dd{x^0} \right) \wedge
    \dd{\phi_1}\wedge \dd{\phi_2}\wedge \dd{\phi_3} \\
    & + 4i
    \frac{z^3}{\rho}\sin r \dd{x^0}\wedge \dd{x^1}\wedge \dd{\theta}
     \wedge \dd{r} \wedge \dd{\xi}\,, \\
    \Phi ={}& 8\eta\, x^0 - 4\eta\, x^1+\log\left[\frac{z^3}{\rho
        \sin^2 r \cos r \sin\xi \cos\xi}\right]\,,
  \end{aligned}
\label{T-HvT}
\end{equation}
where the other components are zero. 

\medskip 

Just like in the first examples that we have discussed, there is a
simple change of coordinates
\begin{align}
  \rho =\tilde{\rho}\,{\rm e}^{-2\eta\, x^1}\,, \qquad z =\tilde{z}\,{\rm e}^{-2\eta\, x^0}\,, 
\end{align}
that diagonalizes the metric:
\begin{equation}
  \begin{aligned}
    \dd{s}^2 ={}& {\rm e}^{-4\eta
      x^0}\tilde{z}^2\left[-(\dd{x^0})^2+(\dd{x^1})^2\right]
    +\frac{ {\rm e}^{4\eta(x^0-x^1)}\dd{\tilde{\rho}}^2+\dd{\tilde{z}}^2}{\tilde{z}^2}
    +{\rm e}^{-4\eta(x^0-x^1)}\frac{\tilde{z}^2\dd{\theta}^2}{\tilde{\rho}^2} \\
    & + \dd{r}^2+\sin^2 r \dd{\xi}^2+ \frac{\dd{\phi_1}^2}{ \sin^2
      r\cos^2\xi} + \frac{\dd{\phi_2}^2}{\sin^2r\sin^2\xi}
    +\frac{ \dd{\phi_3}^2}{\cos^2r}\,,\\
    \mathcal{F}_5 ={}&
    \frac{-4i {\rm e}^{2\eta (x^0-x^1)}}{\tilde{z}^2\sin^2r\cos
      r\sin\xi\cos\xi}\,
    \dd{\tilde{\rho}} \wedge \dd{\tilde{z}} \wedge \dd{\phi_1}
    \wedge \dd{\phi_2}\wedge \dd{\phi_3} \\
    & + 4i
    \frac{ {\rm e}^{-2\eta\,(3x^0-x^1)}\tilde{z}^3}{\tilde{\rho}}\sin r\,
    \dd{x^0}\wedge \dd{x^1}\wedge \dd{\theta} \wedge \dd{r} \wedge \dd{\xi}\,,  \\
    \Phi ={}& 2\eta\,( x^0 -
    x^1)+\log\left[\frac{\tilde{z}^3}{\tilde{\rho} \sin^2 r \cos r
        \sin\xi \cos\xi}\right]\,.
  \end{aligned}
\end{equation}
In this case, however, the linear dependence of the dilaton on the T-dual variables remains.

\medskip 

It may be helpful to recall the result for the case of 3D Schr\"odinger spacetime~\cite{KY-Sch}. 
An affine symmetry algebra is given by a twisted Yangian, which is called an exotic symmetry 
in~\cite{KY-Sch}, and can be mapped to the standard Yangian by undoing the integrable twist. 
Then, however, the target spacetime is not mapped to the undeformed AdS$_3$\,. 
The resulting geometry is described by a dipole-like coordinate system and
 hence it is very close to, but not identical to AdS$_3$\,.

\subsubsection*{The case without the total derivative of $B_2$}

It would be interesting to study also the case without the total derivative term of $B_2$ in (\ref{HvT})\,. 
Then the ``T-dualized'' background is different from the one of  Eq.~(\ref{T-HvT}) 
and the resulting background is given by
\begin{eqnarray}
\dd{s}^2 &=& -(z^2-4\eta^2)\,(\dd{x^0})^2+z^2\,(\dd{x^1})^2
+\frac{(\dd{\rho}+2\eta\rho \dd{x^1})^2}{z^2}+\frac{\dd{z}^2}{z^2-4\eta^2}
+\frac{z^2\dd{\theta}^2}{\rho^2}\no \\
&& + \dd{r}^2+\sin^2 r \dd{\xi}^2+ \frac{\dd{\phi_1}^2}{ \sin^2 r\cos^2\xi} 
+ \frac{\dd{\phi_2}^2}{\sin^2r\sin^2\xi}
+\frac{ \dd{\phi_3}^2}{\cos^2r}\,,\no\\ 
\mathcal{F}_5 &=& 
\frac{-4i}{z^2\sin^2r\cos r\sin\xi\cos\xi}\,
\left( \dd{\rho}+2\eta\, \rho \dd{x^1}\right) \wedge 
\left(-\frac{z^2\dd{z}}{z^2-4\eta^2}+2\eta\, z \dd{x^0} \right) \wedge 
\dd{\phi_1}\wedge \dd{\phi_2}\wedge \dd{\phi_3}\no \\
&& + 4i  
\frac{z^3}{\rho}\sin r\, \left(\dd{x^0}-\frac{2\eta\dd{z}}{z(z^2-4\eta^2)}\right)
\wedge \dd{x^1}\wedge \dd{\theta} \wedge \dd{r} \wedge \dd{\xi}\,, \no \\
\Phi &=& -8\eta\, x^0 - 4\eta\, x^1
+\log\left[\frac{(z^2-4\eta^2)^2}{\rho z \sin^2 r \cos r \sin\xi \cos\xi}\right]\,. 
\label{T-HvTno}
\end{eqnarray}
This background is a solution of the usual type IIB supergravity, 
{and agrees with the one obtained 
in~\cite{HvT} after fixing some typos\footnote{
We would like to thank Ben Hoare and Stijn van Tongeren for clarifying this point.}.} 

\medskip

Now it is natural to ask whether (\ref{T-HvT}) and (\ref{T-HvTno}) are equivalent or not,
 and if so, whether this equivalence holds locally or globally.  
The local equivalence can be shown explicitly by using the coordinate 
transformations\footnote{The authors would like to thank Ben Hoare and Stijn van Tongeren for pointing out
the coordinate transformations.}
\begin{eqnarray}
x^0&\to&-x^0+\frac{1}{4\eta}\log\left[\frac{z^2-4\eta^2}{z^2}\right]\qquad z>2\eta\,,\no \\
x^0&\to&-x^0+\frac{1}{4\eta}\log\left[\frac{4\eta^2-z^2}{z^2}\right]\qquad z<2\eta\,.
\end{eqnarray}
For the case of the global equivalence, some subtleties arise. 
{The background (\ref{T-HvT}) is regular while the one (\ref{T-HvTno}) has a coordinate singularity 
at $z=2\eta$ and so are the coordinate transformations. }
Moreover, two types of time directions have to be introduced. 
Due to these observations, a more involved analysis is necessary in order to argue the global equivalence.

\section{Conclusion and discussion} 
\label{sec:conclusions}

In this paper we have studied the relation between non-Abelian classical $r$-matrices 
and the generalized type IIB equations. We have discussed several examples of 
non-Abelian classical $r$-matrices and derived the corresponding
ten-dimensional  backgrounds by performing a supercoset construction. 

\medskip 

Our main result is that these backgrounds do not satisfy the standard
type IIB equations of motion but a set of generalized equations,
corresponding to a scale-invariant -- as opposed to Weyl-invariant --
worldsheet theory. {This is consistent with the scaling result of \cite{HvT}, 
and the general result of \cite{WT} 
under the assumptions that the deformed string action is the canonical Green-Schwarz string.} 
For some of these examples, we were able to perform a T-duality
transformation leading to a new solution of the standard IIB
equations, where the dilaton depends linearly on some of the T-dual
variables. For some of these examples, we have even been able
to find a change of variables showing that the linear-dilaton
backgrounds are locally T-dual to undeformed \(\ads_{5} \times S^5\). 
This fact makes the classical integrability of these examples manifest.

\medskip 

We have however not been able to find appropriate T-dual frames for all of
the examples. At this point, it is unclear whether this is only due to technical
issues or whether there is an underlying physical reason for this. It would therefore 
be of use to explore more general classical \(r\)-matrices,
as well as lower-dimensional cases, such as deformations of
\(\ads_{2} \times S^2\) or \(\ads_{3} \times S^3\)~\cite{BTW,Lunin}, and 
the relationship between our construction and the \(\lambda\)
deformation of~\cite{lambda1,lambda2,lambda3,lambda4}.

\medskip 

We find that a criterion for distinguishing classes of non-Abelian classical $r$-matrices 
is of importance. A possible criterion is whether a given non-Abelian 
classical $r$-matrix is constructed as a twist of the Drinfeld-Jimbo $r$-matrix or not   
(for details, see Sec.\ 3.1 of~\cite{KMY-Jordanian-typeIIB}). 
This direction would be useful for the classification 
of non-Abelian classical $r$-matrices.  

\medskip 

Gaining a deeper understanding of the underlying algebraic structure giving rise to 
our findings is essential, as it might throw light on the fundamental mathematical underpinnings 
of the gravity/CYBE correspondence.

\subsection*{Acknowledgments}

We are very grateful to Gleb Arutyunov and  
Hideki Kyono for useful discussions. 
We are indebted to Ben Hoare and Stijn van Tongeren for clarifying 
some points and misinterpretations. K.Y.\ would like 
to acknowledge useful discussions which have taken place during the workshop 
``Generalized Geometry \& T-dualities'' at the Simons Center for Geometry and Physics, 
especially valuable comments by A.~A.~Tseytlin and C.~Klimcik.

D.O. and S.R. would like to thank the Department of Physics of Kyoto University for hospitality 
where this work was initiated. 
The work of S.R.\ was supported by the Swiss National Science Foundation
(\textsc{snf}) under grant number PP00P2\_157571/1.
The work of J.S.\ was supported by the Japan Society for the Promotion of Science (JSPS).
The work of K.Y.\ was supported by the Supporting Program for Interaction-based 
Initiative Team Studies (SPIRITS) from Kyoto University and 
by a JSPS Grant-in-Aid for Scientific Research (C) No.\,15K05051.
This work was also supported in part by the JSPS Japan-Russia Research 
Cooperative Program and the JSPS Japan-Hungary Research Cooperative Program.

\end{document}